\documentclass[pre,amsmath,amssymb,twocolumn,footinbib,superscriptaddress]{revtex4}

\usepackage{amsmath,amssymb}
\usepackage[usenames]{color}
\usepackage{amssymb}
\usepackage{dsfont}
\usepackage{grffile}
\usepackage[pdftex]{graphicx}
\usepackage{amsmath, amstext, amssymb, amsfonts, amsxtra}
\usepackage{textcomp}
\usepackage{xspace}
\usepackage{hyperref}
\DeclareSymbolFontAlphabet{\amsmathbb}{AMSb}

\newcommand{\cur}{\mathcal{J}}

\newcommand{\Rec}{\mathcal{R}}

\newcommand{\ave}[1]{\langle #1 \rangle }

\newcommand{\rhop}{\hat{\rho}}

\newcommand{\Hop}{\hat{H}}
\newcommand{\sop}{\hat{\sigma}}
\newcommand{\Dop}{\mathcal{D}}

\newcommand{\im}{{\rm i}}

\newcommand{\An}{\hat{A}_{n}(\omega)}
\newcommand{\Ad}{\hat{A}_{n}^{\dagger}(\omega)}

\newcommand{\w}{\omega}

\newcommand{\Lop}{\mathcal{L}}

\newcommand{\newmod}[1]{{\color{black} #1}}
\newcommand{\kbc}{k_B}

\newcommand{\sutd}{Science and Math Cluster and EPD Pillar, Singapore University of Technology and Design, 8 Somapah Road, 487372 Singapore}
\newcommand{\como}{Center for Nonlinear and Complex Systems, Dipartimento di Scienza e Alta Tecnologia, Universit\`a degli Studi dell'Insubria, via Valleggio 11, 22100 Como, Italy}
\newcommand{\infn}{Istituto Nazionale di Fisica Nucleare, Sezione di Milano, via Celoria 16, 20133 Milano, Italy}
\newcommand{\nest}{NEST, Istituto Nanoscienze-CNR, I-56126 Pisa, Italy}
\newcommand{\brasil}{Departamento de F\'{\i}sica--Instituto de Ci\^{e}ncias Exatas, Universidade Federal de Minas Gerais, CP 702, 30.161-970 Belo Horizonte MG, Brazil}
\newcommand{\riogrande}{International Institute of Physics, Federal University of Rio Grande do Norte, 1613 Natal, Brazil }

\begin{document}

\title{Heat current rectification in segmented $XXZ$ chains}

\author{Vinitha Balachandran}
\affiliation{\sutd}
\author{Giuliano Benenti}
\affiliation{\como}
\affiliation{\infn}
\affiliation{\nest}
\author{Emmanuel Pereira}
\affiliation{\brasil}
\author{Giulio Casati}
\affiliation{\como}
\affiliation{\riogrande}
\author{Dario Poletti}
\affiliation{\sutd}

\begin{abstract}
We study the rectification of heat current in an $XXZ$ chain segmented in two parts.
We model the effect of the environment with Lindblad heat baths. We show that, in our system, rectification is large for strong interactions in half of the chain and if one bath is at cold enough temperature. For the numerically accessible chain lengths, we observe that the rectification increases with the system size. We gain insight in the rectification mechanism by studying two-time correlations in the steady state.
The presence of interactions also induces a strong nonlinear response to the temperature difference, resulting in superlinear and negative differential conductance regimes.
\end{abstract}

\maketitle

\section{Introduction}
Control of nanoscale heat transport can lead to promising applications in thermal waste management. An important step in this direction is devising heat rectifiers (diodes) that allow heat to flow in a preferential direction. Possible pathways to thermal rectification were proposed in classical systems exploiting tailored transport in asymmetric non-linear chains \cite{TerraneoCasati2002, LiCasati2004, Baowen2012, Benenti2016, LeiLi2011, Bagchi2013, KaushikMarathe2018}. In such systems, the rectification mechanism stems from a temperature dependent mismatch in the power spectrum of portions of the nonlinear chain. These theoretical results were confirmed in pioneering experimental investigations of thermal rectification of phonon transport \cite{Chang,Kobayashi}.

Investigations of rectifiers have been done also at the quantum level using other energy carriers such as photons \cite{Ruokola}, electrons \cite{Kuo,Ren,Perez} and spins \cite{Segal2005,Segal,Segal2009,ArracheaAligia2009,LandiKarevski2014,Yan2011,Flores,Chen}. In \cite{Segal,Segal2009}, asymmetric quantum structures and hybrid material junctions are identified as the main ingredients for thermal rectification. From this perspective, spin-boson models (spin chains connected to bosonic baths) have been extensively studied \cite{Yan,LifaLi2009,WerlangValente2014,MotzAnkerhold2018}.
In \cite{MotzAnkerhold2018}, weak anharmonicities were used to generate some heat current rectification. In \cite{WerlangValente2014}, it was shown that two spins coupled via Ising interaction can work as a perfect heat rectifier in the regime of strong spin-spin coupling. However, studies in longer chains observed moderate rectification.
In short, an effective way to construct a well-performing heat diode for larger systems is an open challenge. This quest is of utmost practical relevance, since it is difficult to apply large temperature biases on systems of small size \cite{size_note}.

Recently, it was shown in \cite{BalachandranPoletti2018} that large interactions in a segmented $XXZ$ chain can produce strong spin current rectification when the system is coupled to magnetization baths (and not heat baths). In particular, the authors showed that when the system was biased in one direction a spin current could flow diffusively, but when the system was biased in the opposite direction, and the interactions strong enough, the current was strongly suppressed and the system became an insulator. The mechanism at the basis of such dynamics was recognized to be the emergence of an excitation gap only in reverse bias and for strong enough interactions.
If a tunnelling between the two halves of a segmented chain is possible at no energy cost, then a current is generated. If however a tunnelling between the two halves can occur only after overcoming an energy gap, then the excitation generated at the interface is localized and no current is generated.
The rectification improves with the system size,
and both numerical results and theoretical arguments suggest a perfect spin diode in the thermodynamic limit. It was also noticed, in \cite{BalachandranPoletti2018}, that rectification is significantly stronger when one bath tends to polarize the chain, so that all spins point down (or up).
The  translation of the above results to thermal rectification is not trivial as (i) one should consider heat instead of spin baths and (ii) it is unclear whether the heat rectification effect remains sizeable when the
temperature of the cold heat bath is not close to zero temperature.

In this work, we show that a segmented $XXZ$ chain
can be a well-performing heat diode.
We find that large rectification can be observed when the interaction is strong enough,
even for not excessively cold temperatures (i.e., comparable with
the other energy scales in the model).
We characterize the dependence of rectification on the interactions, interface coupling and baths temperatures, and discuss in which scenarios the rectification is stronger.
For the chain lengths accessible to numerical investigations, we observe that
the rectification increases with the system size.
To gain a deeper insight of the physical mechanism behind thermal rectification, we also analyse the two-time correlations of a tunneling excitation in the middle of the chain. We observe that only in reverse bias the frequency response of the system depends significantly on the magnitude of the interaction.
Finally, we characterize the nonlinear response of the system to temperature differences. In this respect we show that it is possible to observe both superlinear and negative differential conductance, the latter being a key feature for building up a thermal transistor.

This paper is organized as follows: in Sec.\ref{sec:model} we introduce the model, in Sec.\ref{sec:results} we present our results and in Sec.\ref{sec:conclusions} we draw our conclusions.

\section{Model}\label{sec:model}
We study a spin-$1/2$ chain segmented in two parts, described by the $XXZ$ Hamiltonian
\begin{align}\label{ham}
  &\Hop = \Hop_L + \Hop_R + \hat{h}_{N/2}(J_{ N/2},0), \\   
  &\Hop_L = \!\!\sum_{n=1}^{N/2-1} \hat{h}_n(J,\Delta),\;\;\;\; \Hop_R = \!\!\sum_{n=N/2+1}^{N} \hat{h}_n(J,0),\\
  &\hat{h}_n(J,\Delta) = J(\sop_{n}^{x}\sop_{n+1}^{x}+\sop_{n}^{y}\sop_{n+1}^{y}) + \Delta\sop_{n}^{z}\sop_{n+1}^{z},
\end{align}
where $J$ and $\Delta$ are the magnitudes of the $XX$ tunneling and the $ZZ$ coupling, $\sop_{n}^{\alpha}$, with $\alpha=x,y,z$ are the Pauli matrices for the $n$-th spin and $N$ is the (even) total number of sites in the chain. The ratio $\Delta/J$ signals the strength of the interaction and it is also referred to as the anisotropy parameter. $J_{N/2}$ is the magnitude of the coupling between the two half-chains, and, as we will show later, plays an important role in determining the system's response.

The chain is coupled to heat baths at different temperatures at its edges. The effect of the heat baths on the system is modeled by a Lindblad master equation \cite{Lindblad,GoriniSudarshan},
\begin{align}\label{mastereq}
  \frac{d\rhop}{d t}=-\frac{\im}{\hbar}[\Hop,\rhop]+\Dop_1(\rhop)+\Dop_N(\rhop)=\Lop(\rhop),
\end{align}
where $\rhop$ is the density operator of the system, $\hbar$ is the (reduced) Planck constant, and $\Dop_1$ ($\Dop_N$) is the dissipator due to the coupling of the first (last) site with the left (right) bath.
Each dissipator is given by \emph{global} operators, that is operators that, in general, act on the full system \citep{PetruccioneBreuer2002}. More specifically,
\begin{align}
\Dop_n(\rhop)=&\sum_{\w>0}J(\w) \left\{ \left[1+n_n(\w)\right] \left[\An\rhop\Ad \right.\right. \nonumber \\
& \left.- 1/2 \left(\Ad\An\rhop + \rhop\Ad\An  \right)  \right]    \nonumber \\
&+n_n(\w) \left[\Ad\rhop\An \right.   \nonumber \\
& \left. \left. - 1/2 \left(\An\Ad\rhop + \rhop\An\Ad\right)  \right]\right\}.
\end{align}
Here, $\w=\epsilon_{j}-\epsilon_{i}$ is the energy difference of two eigenstates, $|\epsilon_i\rangle$ and $|\epsilon_j\rangle$, of $\hat{H}$, $n_n(\w)=[\exp(\hbar\w/\kbc T_{n})-1]^{-1}$ is the Bose-Einstein distribution characterizing the heat baths ($n=1$ or $N$) and $\kbc$ is the Boltzmann constant. The Lindblad operators $\An=\sum_{\w} |\epsilon_i \rangle|\langle \epsilon_i |\sop^{x}_{n}|\epsilon_j\rangle\langle \epsilon_j|$ describe the transitions induced by the bath. In the following we use an Ohmic bath, with the bath spectral function $J(\w)=\gamma \w$. This master equation can be justified in weak coupling limit and it has been shown that it can predict accurately the heat current exchanged between the system and the baths \cite{WichterichMichel2007, Alicki1979, LevyKosloff2014}. To lighten the notations, henceforth we will work in units in which $J=k_B=\hbar=1$.

The heat current $\cur$ is calculated from the continuity equation,
\begin{align}
\frac{d\ave{\Hop}}{dt}=\cur_L-\cur_R.
\end{align}
It follows from the master equation in Eq. (\ref{mastereq}) that
 \begin{align}
\frac{d\ave{\Hop}}{dt}=\mathrm{Tr}\big[\Hop\Dop_1(\rhop)\big]+\mathrm{Tr}\big[\Hop\Dop_N(\rhop)\big].
\end{align}
In the steady state, $\frac{d\ave{\Hop}}{dt}=0$.
Thus, the steady state heat current is
 \begin{align}
\cur=&\cur_{L}=\cur_R=\mathrm{Tr}\big[\Hop\Dop_1(\rhop_s)\big]=-\mathrm{Tr}\big[\Hop\Dop_N(\rhop_s)\big], \label{eq:currentdef}
\end{align}
where $\rhop_s=\rhop(t=\infty)$ is the steady state.
We refer to {\it forward bias} the case in which the hot bath is coupled to the first (left-most) site and the cold one to the last (right-most) site, while we use {\it reverse bias} for the opposite case.
The magnitude of the rectification is signalled by the rectification coefficient, which is the ratio between the current in forward bias $\cur_{f}$ and that in reverse bias $\cur_{r}$:
 \begin{equation}\label{rect2}
  \Rec=-\frac{\cur_{f}}{\cur_{r}}.
 \end{equation}
The rectification coefficient $\Rec$ is much larger, or much smaller, than $1$ for good rectifiers \cite{fn1}, while $\Rec=1$ for no rectification \cite{indgamma}.

\section{Results}\label{sec:results}

We now study how the magnitude of the anisotropy $\Delta$, the temperature $T_C$ ($T_H$) of the cold (hot) bath, and the interface coupling $J_{ N/2}$ affect the rectification obtained in the system.
The steady state is calculated from the stationary solution of the master equation (\ref{mastereq}),
 \begin{equation}
 \Lop(\rhop_s)=0,
 \end{equation}
i.e., the zero$-$th mode of the Lindblad super operator $\Lop$ which in our case is unique. The memory cost of finding the eigenvector for the zero$-$th eigenvalue of $\Lop$ using exact diagonalization is $4^{2N}$. Thus, it is highly non-trival to go beyond $N>8$. So for larger chains, we employ Runge-Kutta method to numerically integrate the master equation and find the steady state. Although the memory cost is reduced to $2^{2 N}$, long integration times are required for convergence to the steady state, and this limits the accessible chain lengths
to $N\approx 10$.

\subsection{Role of interaction}
\begin{figure}
\includegraphics[width=\columnwidth]{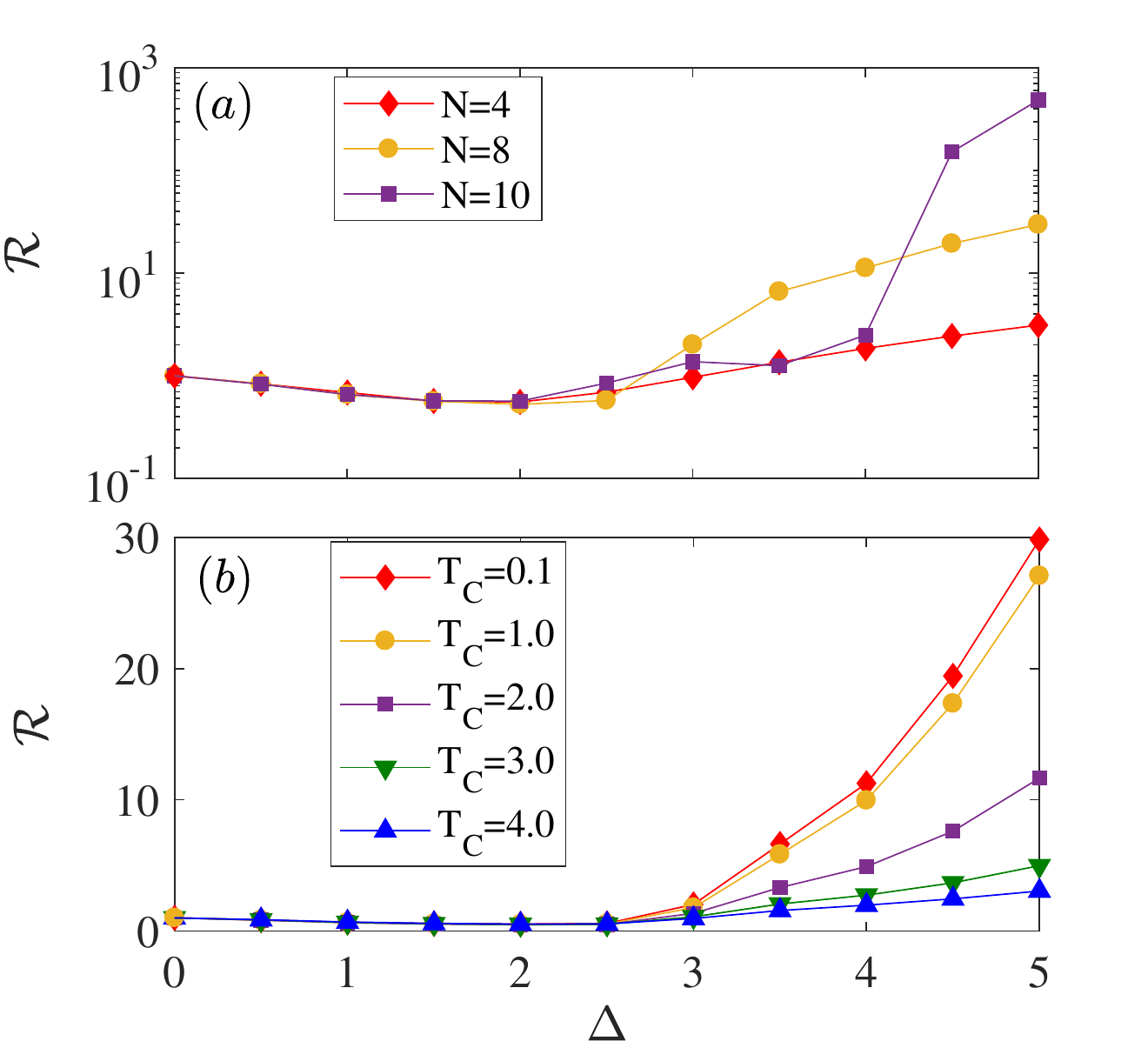}
\caption{{\bf Role of interaction. } (a) Rectification coefficient $\Rec$ as a function of anisotropy parameter $\Delta$  for different chain lengths $N$. Cold bath temperature is $T_{C}=0.1$ and hot bath temperature is $T_{H}=10+T_{C}$. (b) Dependence of rectification coefficient $\Rec$ on anisotropy parameter $\Delta$  with different cold bath temperature $T_{C}$ for a chain of length $N=8$ and $T_{H}=10+T_{C}$. In both panels we used $J_{N/2}=1$.
}
\label{fig:Fig1}
\end{figure}
First, we investigate the effect of interactions on the rectification. Fig.\ref{fig:Fig1} shows the variation of rectification $\Rec$ with the interaction strength $\Delta$. Similar to the spin current in \cite{BalachandranPoletti2018}, rectification increases significantly when $\Delta \gtrsim 2$. In Fig.\ref{fig:Fig1}(a) in particular, we show how the rectification changes with the chain length $N$. For the system sizes analyzed, we find that significant rectification can be observed even for small $N$. Moreover we observe an increase of the rectification for larger $N$. We are however unable to directly explore large system sizes and cannot predict on analytical grounds what would be the behavior in the thermodynamic limit.

However we can conjecture what would happen in the thermodynamic limit by reverting to results for spin currents and spin baths as in \cite{BalachandranPoletti2018}. In that system, in fact, we observed that when the fully polarizing bath was not pushing the system exactly in the state $|\downarrow\downarrow\dots\downarrow\rangle$, not only the rectification was reduced, but $\Rec$ would decrease with the system size. Hence, we expect that also in our set-up, as soon as $T_C>0$, the system will never be perfectly insulating in reverse bias, hence the temperature bias per unit-length will be reduced as the system size increases, thus bringing the system towards linear response where rectification is not possible. This argument implies that for a given bath temperature there is an optimal system size for heat current rectification.

In Fig.\ref{fig:Fig1}(b) we shift our focus on the importance of the temperature of the cold bath $T_C$. Here we show that going to low values of $T_C$ can induce a significant increase in the observed rectification. This is consistent with \cite{BalachandranPoletti2018} in which, when one of the baths was trying to fully polarize the chain, the rectification was stronger.

\subsection{Role of $\Delta T$}
\begin{figure}
\includegraphics[width=\columnwidth]{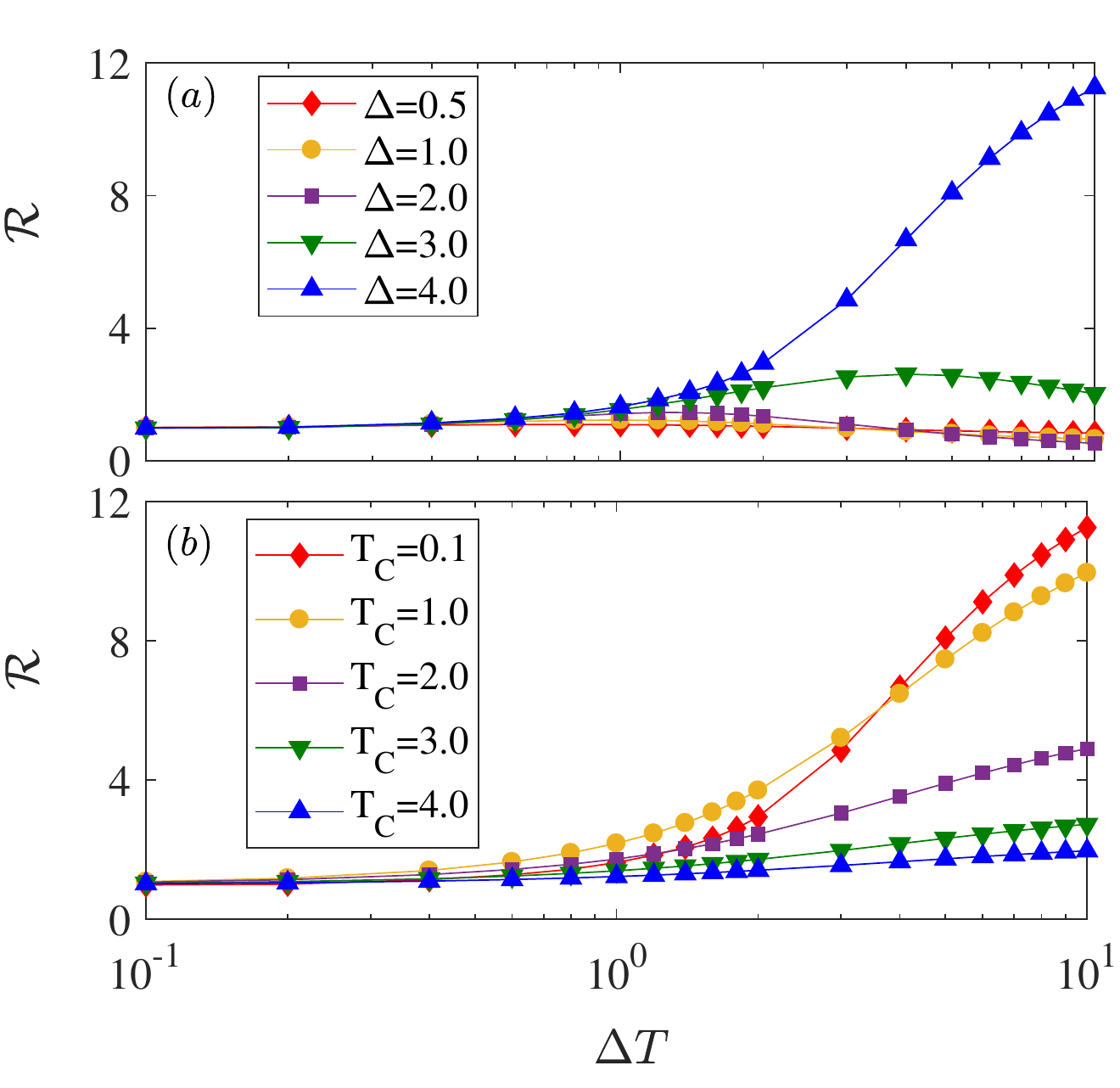}
\caption{{\bf Role of $\Delta T$. }(a) Variation of rectification coefficient $\Rec$ with temperature bias $\Delta T$ for different anisotropy parameter $\Delta$. Here, cold bath temperature is $T_{C}=0.1$. (b) Dependence of rectification coefficient $\Rec$ on temperature bias $\Delta T$  for different cold bath temperature $T_{C}$. Here the anisotropy $\Delta=4$. In both panels the chain length $N=8$, $T_{H}=T_{C}+ \Delta T$ and $J_{N/2}=1$.
}
\label{fig:Fig2}
\end{figure}
In this section, we will detail the importance of the temperature bias in the rectification mechanism. An increase in $\Delta T= T_H-T_C$ allows a strong nonlinear response and hence the possibility of strong rectification. In Fig.\ref{fig:Fig2}(a) we plot the rectification versus $\Delta T$ for different magnitudes of the anisotropy, while keeping the cold temperature at $T_C=0.1$. For small $\Delta T$, the current increases both in the forward and reverse biases, but in a very similar manner and the rectification remains always $\mathcal{R} \approx 1$. When $\Delta T \gtrsim 1$ strong rectification appears, however, it is only for large anisotropy $\Delta$ that $\mathcal{R}$ become particularly prominent.
We also observe a decrease of rectification with larger $\Delta T$,
and eventually a large bias favours reverse current over the forward current, thus resulting in a rectification coefficient $\Rec<1$.
Since we numerically observed that the value of $\Delta T$ where
rectification reversal takes place increases with the system size, we
interpret this phenomenon as a proximity effect related to the small
chain length (see Appendix A for more details).

In Fig.\ref{fig:Fig2}(b) we instead show the rectification versus the temperature difference $\Delta T$ but for different values of the cold temperature $T_C$, while keeping $\Delta=4$. Fig.\ref{fig:Fig2}(b) confirms that, even for large anisotropy, it is only when the cold temperature $T_C$ is low enough that one observes strong rectification.

\subsection{Role of interface tunneling}

\begin{figure}
\includegraphics[width=\columnwidth]{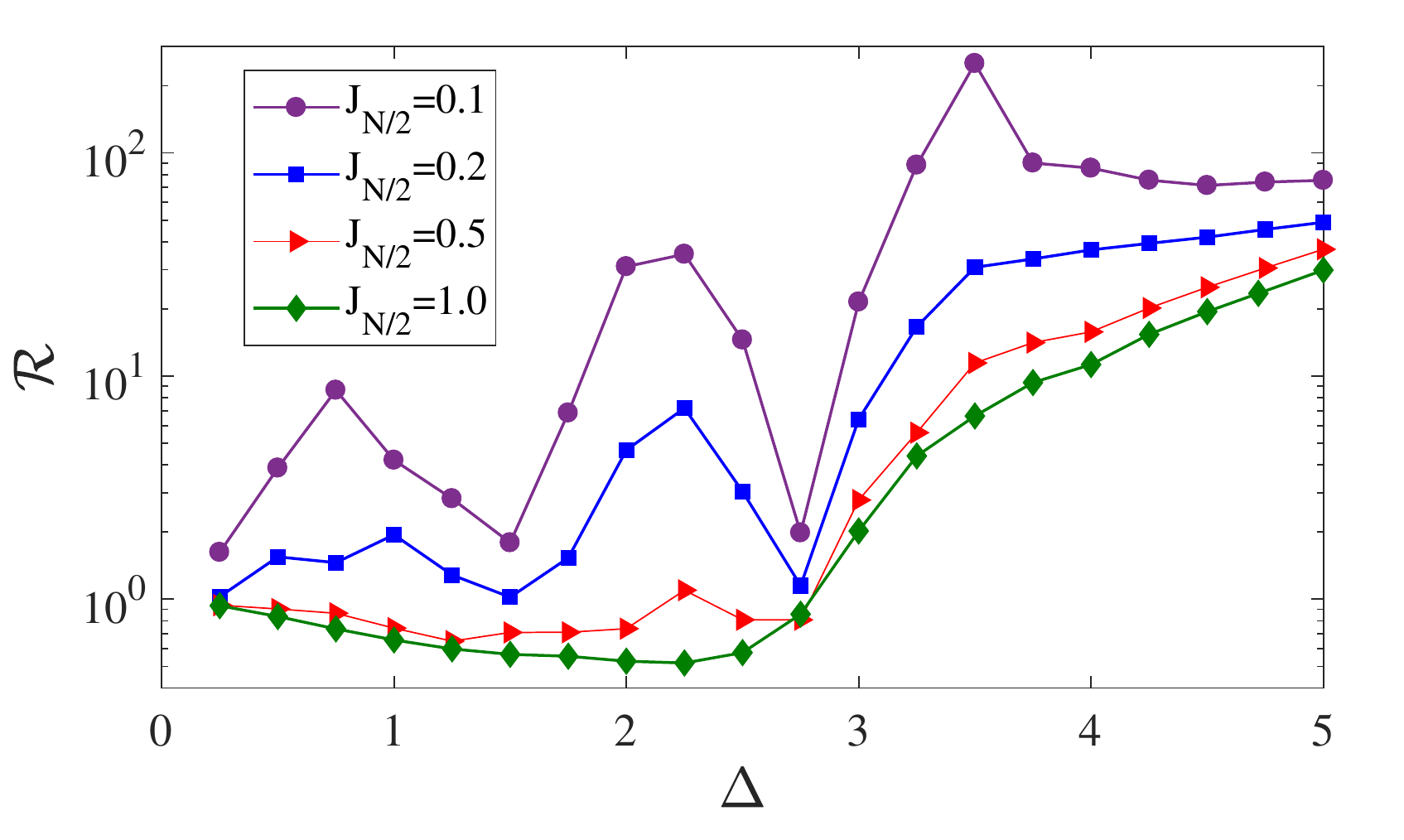}
\caption{{\bf Interface tunneling.} Rectification ratio versus anisotropy $\Delta$ for different magnitudes of the tunneling at the interface between the interacting and the non-interacting halves of the chain (purple circles for $J_{N/2}=0.1$, blue squares for $J_{N/2}=0.2$, red triangles for $J_{N/2}=0.5$ and green diamonds for $J_{N/2}=1.0$). The smaller the interface tunneling $J_{N/2}$ the stronger the rectification effect, however larger are the resonance effects due to finite size. Other parameters are $N=8$, $T_C=0.1$ and $T_H=10$. }
\label{fig:Fig3}
\end{figure}

The tunnelling at the interface $J_{N/2}$ plays a very important role in determining  the amount of current in the system and the magnitude of the rectification. Smaller $J_{N/2}$ implies smaller currents, however it also allows to recognize more clearly the differences in the spectrum of the halves of the spin chain, by separating/distinguishing them more strongly. This results in stronger rectifications for smaller interface tunnelling $J_{N/2}$, but also in sharper responses close to resonances, where there is no energy gap for the propagation of a magnon excitation created at the interface between the two halves of the segmented chain, and therefore  $\Rec\approx 1$ \cite{BalachandranPoletti2018}. This is clearly depicted in Fig.\ref{fig:Fig3} in which we show $\Rec$ versus the anisotropy $\Delta$ for different values of the interface tunneling $J_{N/2}$.

\subsection{Frequency response}

\begin{figure}
\includegraphics[width=\columnwidth]{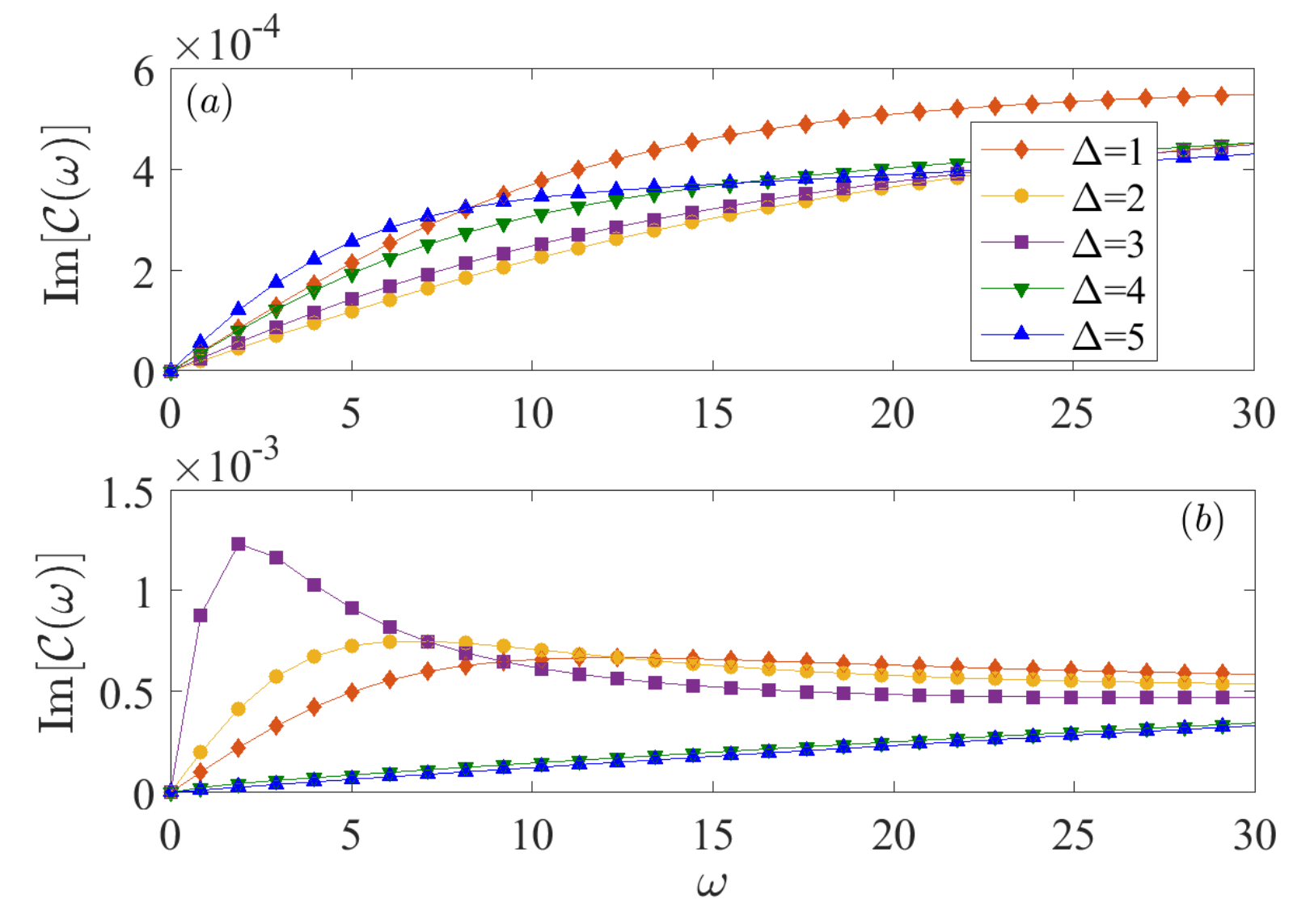}
\caption{{\bf Frequency response.} ${\rm Im}[\mathcal{C}(\omega)]$, i.e. the imaginary part of the Fourier transform of the two-time correlation $C(t)$ in Eq.(\ref{eq:correlation}) as a function of the frequency $\omega$. We consider different values of the anisotropy (red diamonds for $\Delta=1$, yellow circles for $\Delta=2$, purple squares for $\Delta=3$, green downward triangles for $\Delta=4$, and blue upward triangles for $\Delta=5$) for forward, panel (a), and reverse bias, panel (b). Other parameters are $N=8$, $T_C=1$, $T_H=100$, $\gamma=1$ and $J_{N/2}=1$. }
\label{fig:Fig4}
\end{figure}

In order to characterize further the rectification system here studied, we consider the frequency response of a tunnelling excitation at the interface. A tunnelling excitation corresponds to a spin exchange between two nearest neighbour spins.
More precisely, we consider the two-time correlation function
\begin{align}
C(t) = \mathrm{Tr}\left[  \sigma^-_{N/2}\sigma^+_{N/2+1} \; e^{\mathcal{L}t} \left(\sigma^+_{N/2}\sigma^-_{N/2+1} \rhop_s \right)\right]. \label{eq:correlation}
\end{align}
This correlation function corresponds to imposing a tunneling on the steady state, evolving it for a time $t$ with the Lindbladian $\mathcal{L}$ and then tunneling back. The Fourier transform of this two-time correlation, $\mathcal{C}(\omega)$, gives the frequency response of the system. We show the dependence of the imaginary part of ${\rm Im}[\mathcal{C}(\omega)]$ on the frequency $\omega$ for different values of the anisotropy $\Delta$ in Fig.\ref{fig:Fig4}. In forward bias, Fig.\ref{fig:Fig4}(a), the two-time correlation is almost unaffected by a change in the anisotropy $\Delta$ \cite{fn3}. In reverse bias instead, for small values of $\Delta$ we observe that there is a sizeable response at all frequencies, including low frequencies. However, for large enough anisotropies, the low frequency response is significantly suppressed, and only at large enough frequencies (i.e., energies in the system), it is possible to have a sizeable response. This is in agreement with our previous work \cite{BalachandranPoletti2018}, where we showed that in reverse bias there is an excitation gap which hinders spin current, while in forward bias no excitation gap is present. See Appendix B for further insights.

\subsection{Nonlinear response and negative differential conductance}
\begin{figure}
\includegraphics[width=\columnwidth]{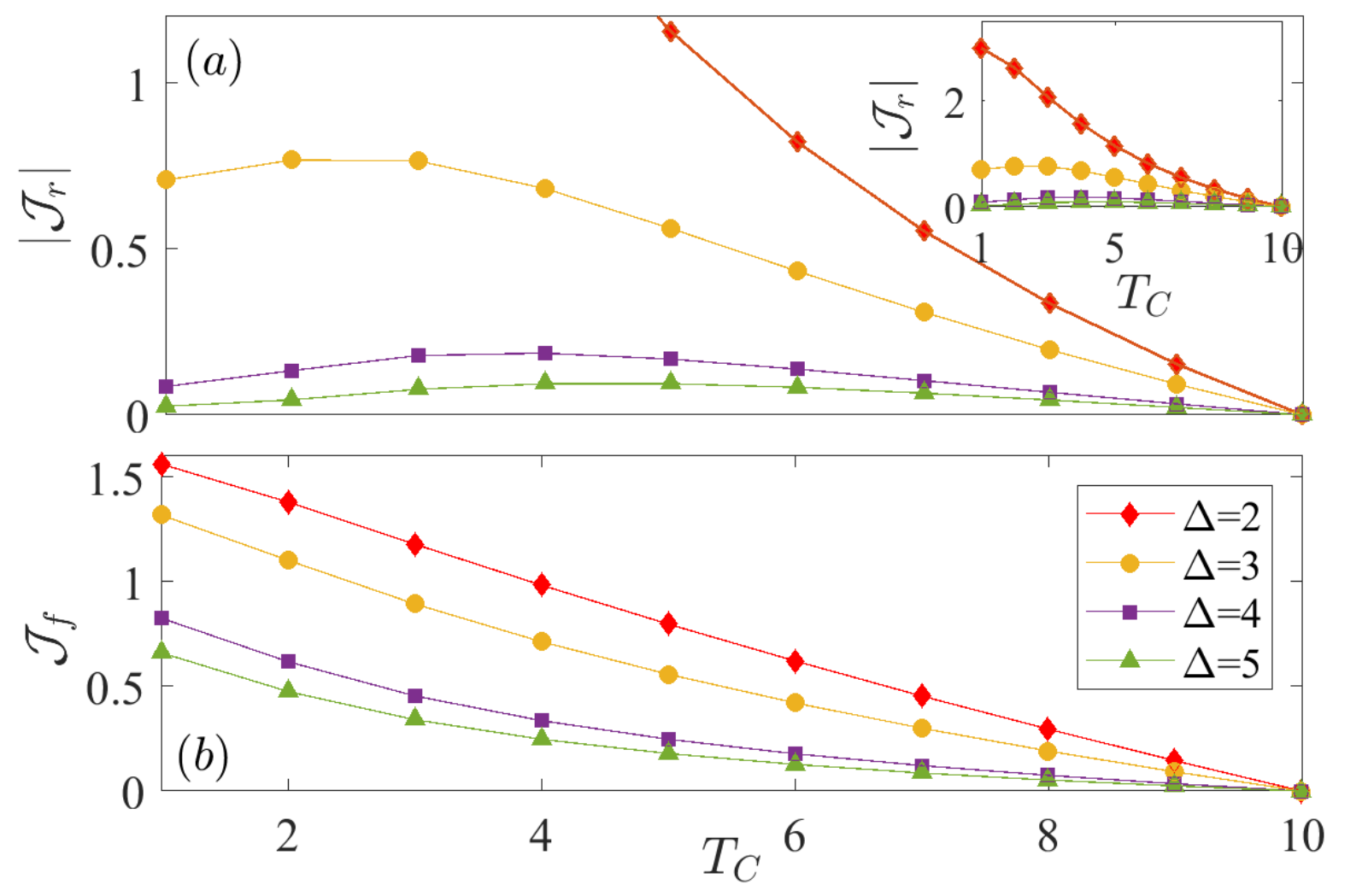}
\caption{{\bf Negative differential conductance.} Reverse, panel (a), and forward bias current, panel (b), vs $T_C$ for different anisotropies $\Delta=5$ (green triangles), $\Delta=4$ (purple squares), $\Delta=3$ (yellow circles) and $\Delta=2$ (red diamonds). The hot bath temperature is set to $T_H=10$. Other parameters are $N=8$, $\gamma=1$ and $J_{N/2}=1$. In the inset of panel (a), the range for the $y$-axis is modified, so that for $\Delta=2$ the monotonic behavior of the heat current can be seen down to low values of $T_C$.}
\label{fig:Fig5}
\end{figure}

Due to the presence of strong interactions, our system offers the opportunity to have nonlinear responses and, in particular, negative differential conductance (see \cite{BenentiRossini2009, BenentiZnidaric2009, GuoPoletti2017} in the context of spins and hard-core bosons transport). Negative differential conductance, i.e. the counter-intuitive phenomenon of decreasing the heat flow despite an increase of thermal bias plays a vital role in the performance of various thermal devices, such as the thermal transistors \cite{LiCasati2006} and thermal memories \cite{WangLi2008}.
It is a very important subject of thermal transport research, and targeting the future possibility of building experimental devices, we want to stress that it is present in our proposal of thermal diode.
In order to show this, we consider again a chain with $N=8$ spins and the first four of them interacting with anisotropy $\Delta$. Current in the system vs the cold temperature is plotted in Fig.\ref{fig:Fig5}(a,b), respectively for reverse and forward biases. We first consider the reverse bias where the first site is set to the cold temperature $T_C$ and the last site to the hot temperature $T_H=10$. For $T_C=10$, the cold bath is at the same temperature as the hot one, so the current is clearly zero. As the cold temperature decreases the current increases until $T_C$ is cold enough and, after that, the current decreases despite the increase in temperature difference $T_H-T_C$. The change of derivative of the current vs temperature curve occurs at higher temperatures for larger anisotropies. At low anisotropy $\Delta$, the current increases super-linearly for small $\Delta T$, i.e. the current increases more than linearly with the change in bias, however, it may show negative differential conductance only for $T_C$ cold enough (e.g. for $\Delta=3$) or not at all (e.g. for $\Delta=2$, not shown in the main panel but highlighted in the inset). In forward bias instead, no negative differential conductance is observed, on the contrary, for larger anisotropy we observe superlinear conductance, as clearly shown in Fig.\ref{fig:Fig5}(b).\\

\section{Conclusions}\label{sec:conclusions}
Motivated by our findings in \cite{BalachandranPoletti2018}, in which it was shown that a segmented $XXZ$ chain can become a perfect diode for spin currents, we investigate the heat current rectification in the same setup. Using global Lindblad heat baths, we show that a segmented chain works as a good diode provided one half of the chain is strongly interacting and the cold bath temperature is low enough. We have shown that the spectral response is significantly suppressed at low frequencies when the anisotropy is large enough. Our simulations show that, up to the numerically accessible chain lengths, the rectification improves with the system size. Unlike the diode for spin currents in \cite{BalachandranPoletti2018}, we find that the heat diode can work over a wide range of bath parameters, although the peak performance of the latter is, in general, worse.

The presence of strong interactions can, in general, induce a nonlinear response to an external bias. Here we show that, depending on the magnitude of the anisotropy and whether the system is in forward or reverse bias, both superlinear and negative differential conductance can emerge, the latter being a key ingredient for building up a thermal transistor.

The presence of a nonlinear response and of strong rectification can be used to control the heat flow and could have applications also in energy conversion from heat to work \cite{BenentiWhitney2017, XuPoletti2018}.

{\bf Acknowledgments}: D.P. and V. B. acknowledge support from the Singapore Ministry of Education, Singapore Academic Research Fund Tier-II (project  MOE2016-T2-1-065). E.P. was partially supported by CNPq (Brazil). G.B. acknowledges the financial support of the INFN through the project “QUANTUM”.


\clearpage


\begin{appendix}
\begin{figure}
\includegraphics[width=\columnwidth]{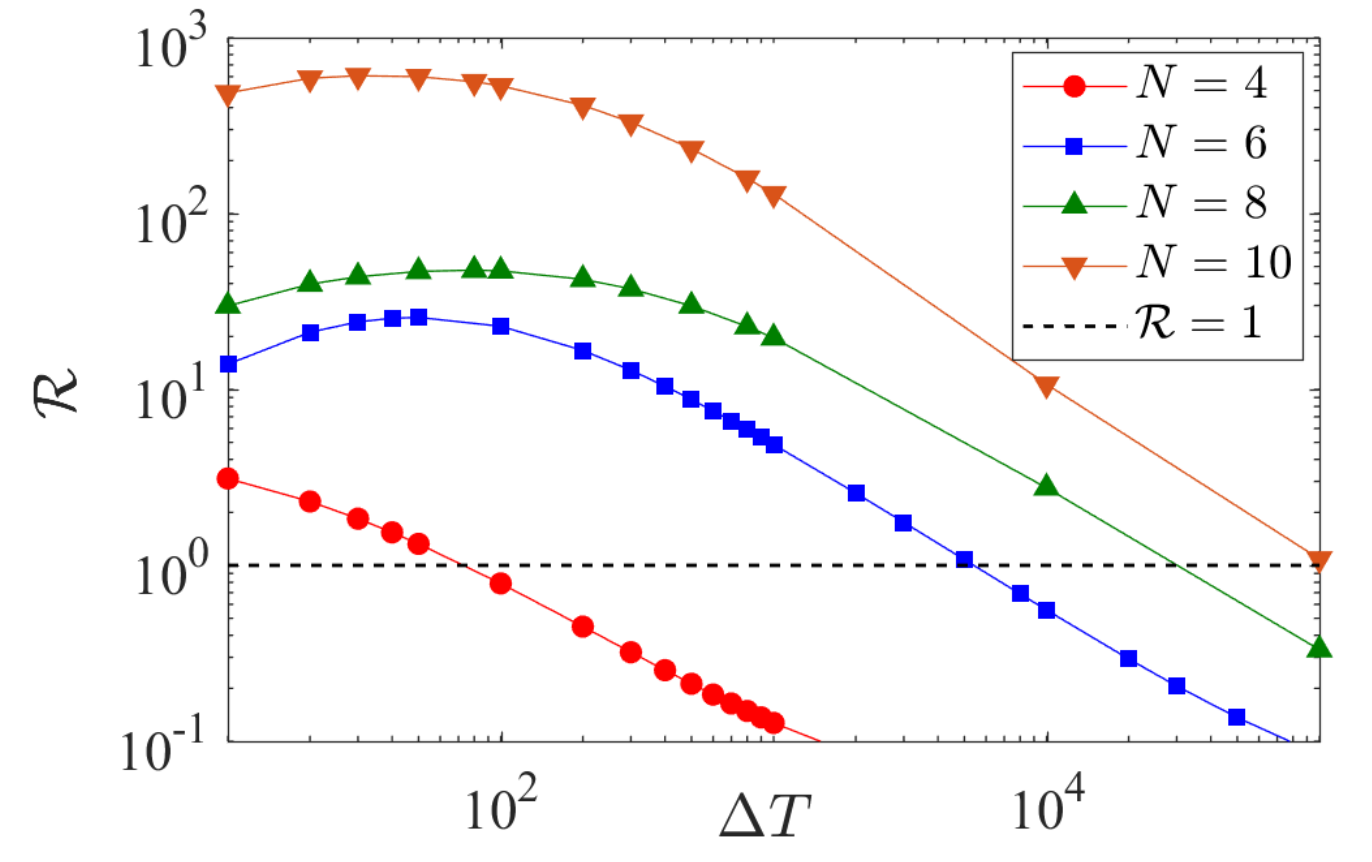}
\caption{{\bf Scaling of rectification reversal. }Variation of rectification coefficient $\Rec$ with temperature bias $\Delta T$ for different chain lengths $N$.  Dotted line corresponds to $\Rec=1. $ Here, the anisotropy is $\Delta=5$, cold bath temperature $T_{C}=0.1$, hot bath temperature $T_{H}=T_{C}+ \Delta T$ and $J_{N/2}=1$.
}
\label{fig:FigS1}
\end{figure}
\section{Reversal of rectification} \label{reversal}
To demonstrate the proximity effect, we study the rectification for different temperature bias $\Delta T$ at different chain lengths $N$ in Fig. \ref{fig:FigS1}. For a chain of spins $N=4$ with anisotropy $\Delta=5$, the rectification $R$ is below $1$ when $\Delta T=74$ whereas for $N=6$, it is after $\Delta T=5000$. The temperature required to reverse the rectification increases to values larger than $10^4$ for $N=8$ and $10^5$ for $N=10$.
From our numerical data we can conclude that, for a longer chain, the increase in the reverse current takes place at larger bias.

\begin{figure}
\includegraphics[width=\columnwidth]{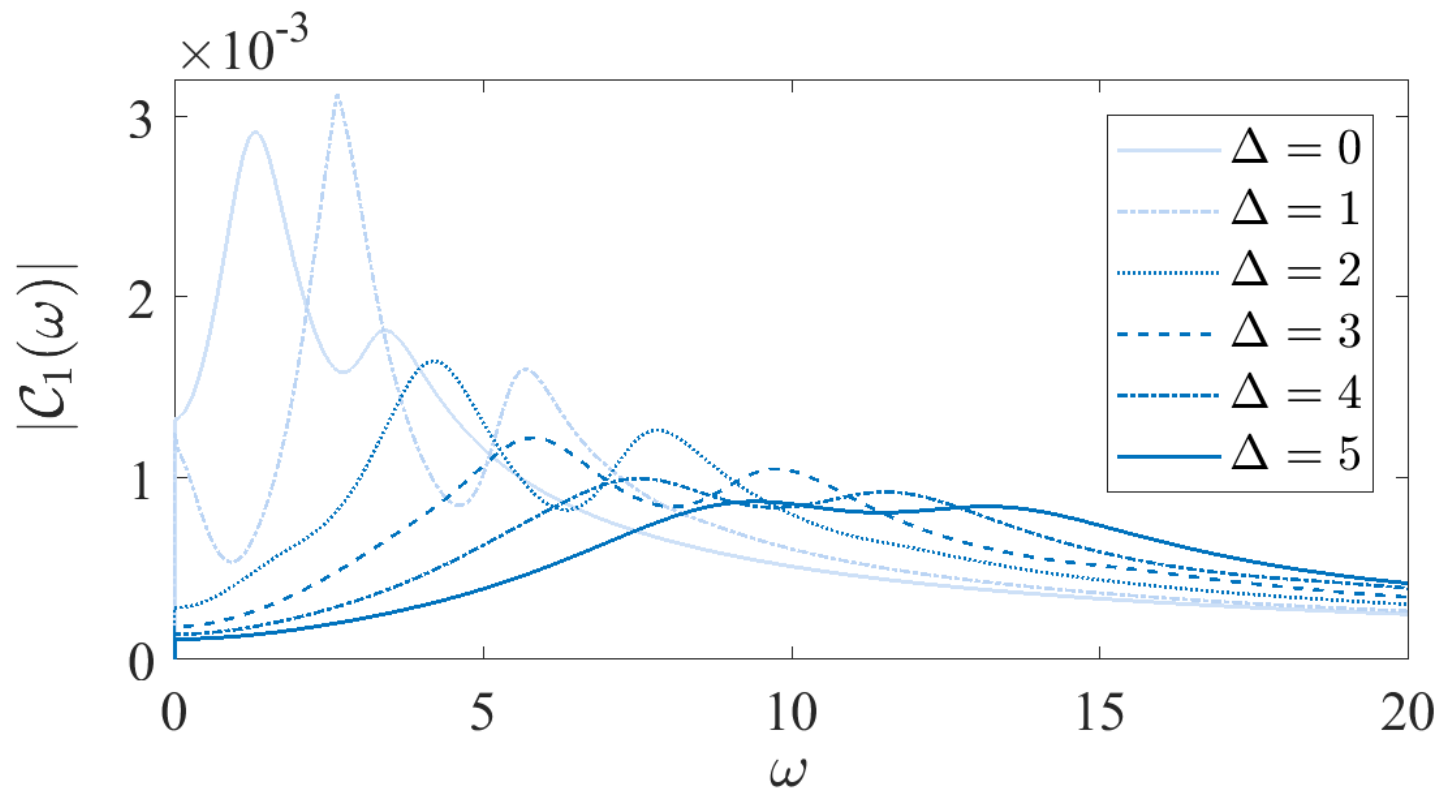}
\caption{{\bf Frequency response for a half-chain.} $|\mathcal{C}_{1}(\omega)|$, i.e. the absolute value of the Fourier transform of the two-time correlation $C_1(t)$ as a function of the frequency $\omega$ with $T=1$, for different anistropies $\Delta$. The length of the chain is $N=4$ and $\gamma=1$. }
\label{fig:FigS2}
\end{figure}

\section{Spectral response for two-time correlation} \label{twotime}
Here, we further show the suppression of the low frequency response in the reverse bias at larger anisotropy. For this, we take the half-chain with uniform anisotropy $\Delta$ at a temperature $T$. The corresponding state of the chain is \newmod{$[\rho_s]_{i,j}=e^{- E_i/T}\delta_{i,j}/\mathcal{Z}$}, where $E_i$ are the eigenenergies of the half chain and $ \mathcal{Z}=\sum_{i}e^{-E_i/T}$. Next, we attach this chain to a single thermal bath with temperature $T$ and calculate the two-time correlation $C_{1}(t) = \mathrm{Tr}\left[  \sigma^+_{1} e^{\mathcal{L}t} \left(\sigma^-_{1} \rho_s \right)\right]$ and its Fourier transform $\mathcal{C}_{1}(\omega)$. The frequency response of the absolute value of $\mathcal{C}_{1}(\omega)$ for different anisotropies is plotted in Fig. \ref{fig:FigS2} at temperatures $T=1$. This figure shows that, at low temperature, the zero frequency response of the chain is highly suppressed when the anisotropy is large. This is a clear indication of opening of an excitation gap, implying small reverse current and large rectification for large anisotropy.

\end{appendix}


\begin{thebibliography}{100}
\bibitem{fn1} When $\Rec$ is much smaller than one, there is a strong rectification but in the opposite direction.
\end{thebibliography}
\end{document}